\documentclass[lettersize,journal]{IEEEtran}
\usepackage{cite}
\usepackage{multicol,lipsum}
\usepackage{amsmath,amssymb,amsfonts}
\usepackage{algorithmic}
\usepackage{graphicx}
\usepackage{textcomp}
\usepackage{xcolor}
\usepackage{booktabs}
 \usepackage{hyperref}
\usepackage{enumerate}
\usepackage{algorithm}
\usepackage{multirow}
\usepackage{standalone}
\usepackage{tikz}
\usetikzlibrary{positioning}
\usetikzlibrary{decorations.markings}

\usepackage{geometry}
\newgeometry{a4paper,scale=0.88}
  \usepackage{enumerate}
\usepackage{subcaption}
 \usepackage{setspace}
\usepackage{verbatim}
   \usepackage{tabularx}
 \usepackage{makecell}

 \usepackage{hyperref}

\begin{document}

\title{ Phase Noise Resilient  Codebook Design for Sparse Code Multiple Access  }

\author{Haibo Liu, Qu Luo,~\IEEEmembership{Member,~IEEE,}
        Zilong Liu,~\IEEEmembership{Senior Member,~IEEE,}\\
          Shan Luo,  
        Pei Xiao,~\IEEEmembership{Senior Member,~IEEE}, 
        and Xiaojun Yuan,~\IEEEmembership{Senior Member,~IEEE.} 
\thanks{
Haibo Liu and  Xiaojun Yuan are with the National Key Laboratory of Wireless Communications, University of Electronic Science and Technology of China (UESTC), China (e-mail: liu\_hb@std.uestc.edu.cn,  xjyuan@uestc.edu.cn).

Qu  Luo  and  Pei  Xiao  are  with  5G \& 6G  Innovation Centre, University of Surrey, UK (email: \{q.u.luo,   p.xiao\}@surrey.ac.uk). 

Zilong   Liu   is   with   the   School   of   Computer   Science   and   Electronics   Engineering,   University   of   Essex,   UK (email:   zilong.liu@essex.ac.uk). 

Shan Luo is with the School of Aeronautics and Astronautics, UESTC , Chengdu, China (e-mail: luoshan@uestc.edu.cn). 
 }
 }


\maketitle

 \vspace{-2em}
 
\begin{abstract}
Sparse code multiple access (SCMA) is a promising technique for future machine type communication systems due to its superior spectral efficiency and capability for supporting  massive connectivity.
This paper proposes a novel class of sparse codebooks  to improve the error rate  performance of SCMA in the presence of  phase noise (PN). Specifically, we first analyze the error rate  performance  of SCMA impaired by  looking into the pair-wise error probability. Then, a novel codebook design metric, called minimum PN metric (MPNM), is proposed.  In addition, to design PN resilient codebooks,  we propose a   novel  pulse-amplitude modulation (PAM)-based low projection mother constellation (LP-MC), called LP-PAM. The codebooks for different users are obtained by rotating and scaling the  MC, where the phase rotation angles and scaling factors for different users are  optimized by maximizing the proposed MPNM. Numerical results show that the proposed  PNCBs have larger MPNM values and achieve improved error rate performance than the-state-of-the-art codebooks.
\end{abstract}


\vspace{-1em}
\section{Introduction}
\IEEEPARstart{W}{ith} the  widespread proliferation of
Internet-of-Things (IoT) across every corner of this globe, new multiple access techniques are required to support improved spectrum efficiency and provide massive connectivity \cite{Sparse}. 
Non-orthogonal multiple access (NOMA) has emerged as a promising  candidate  to meet these requirements in recent years \cite{Sparse,SSDSCMA}. Compared to the orthogonal multiple access techniques which allocate  users with orthogonal time/frequency resources, NOMA allows several times more users  to share the same radio resources simultaneously. This paper is concerned with sparse  code multiple access (SCMA) which is a representative code-domain NOMA  scheme  \cite{SSDSCMA, PAPRSCMA,CBDesign,ZhangUDGG,UplinkRician,LPSCMA}. 
Over the past few years,  there have been a number of research attempts  on SCMA codebook design \cite{{PAPRSCMA,CBDesign,ZhangUDGG,UplinkRician,LPSCMA,POSCMA}}. By looking
into the pair-wise error probability (PEP) over Gaussian and Rayleigh fading channels,
it is desirable to maximize the minimum Euclidean distance
(MED) and minimum product distance (MPD) of a mother constellation (MC) or a codebook. Following this design guideline, an SCMA codebook design approach   based on uniquely decomposable constellation group was proposed in \cite{ZhangUDGG} by maximizing the MED. \cite{POSCMA} proposed a class of power imbalanced codebook by maximizing the MED while keeping the MPD as large as possible. The SCMA codebook design has also been studied in Rician fading channels \cite{UplinkRician,LPSCMA}. Specifically,  \cite{UplinkRician} and \cite{LPSCMA} have proposed a novel class codebooks for uplink and downlink Rician fading channels, respectively.

Generally speaking, the existing SCMA codebooks are mostly optimized with respect to certain wireless channel conditions but with very few considerations on the hardware impairments. The latter is particularly important for an SCMA based IoT system supporting the information exchanges over  low-cost and low-end communication devices.  {However,
existing metrics for SCMA optimization, such as MED or PEP, do not adequately capture
phase noise (PN) effect, necessitating the development of a new metric tailored to this impairment. Among many others, this paper aims to develop novel sparse codebooks which are resilient to PN\cite{likelyhood_kam}.}  
By looking into the PEP, we analyse the theoretical performance of SCMA system in the presence of PN and propose a new distance metric, called minimum PN metric (MPNM).  We then propose a novel pulse-amplitude modulation (PAM)-based low projection mother constellation (LP-MC) for PN resilience. Finally, a novel class of PN resilient  codebooks is obtained by maximizing the proposed MPNM based on the proposed LP-MC. 

The rest of the paper is organized as follows: Section \ref{SystC} introduces SCMA communication model impaired by PN. The error rate performance  and the codebook design criteria of SCMA impaired by PN are introduced in Section \ref{PEPS}. Then, Section  \ref{CBd} presents the proposed codebook by maximizing the MPNM.   Numerical results are provided in Section \ref{Sim}.  Section \ref{conclu}  concludes the paper.

\IEEEPARstart{}{} 

\vspace{-2em}
\section{System Model}
\label{SystC}
\subsection{ SCMA Communication Model Impaired by Phase Noise}
\begin{figure}[tbp]
    \centering    \includegraphics[width=0.75\linewidth]{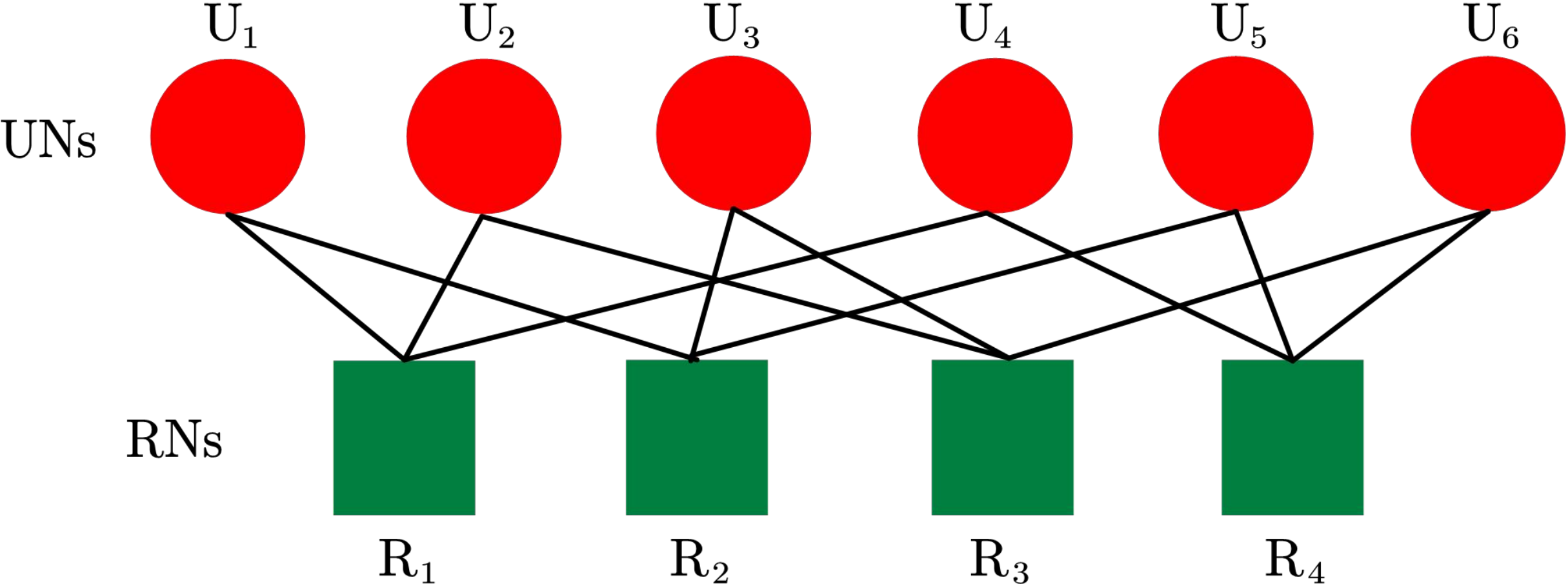}
    \caption{An example of SCMA factor graph ($K=4$, $J=6$).}
    \label{Factor}
    \vspace{-1em}
\end{figure}

\begin{figure}[tbp]
\centering
\subfloat[Chen's codebook\cite{Chen}]{
\begin{minipage}[t]{0.42\linewidth}
\centering
\includegraphics[width=1.3in]{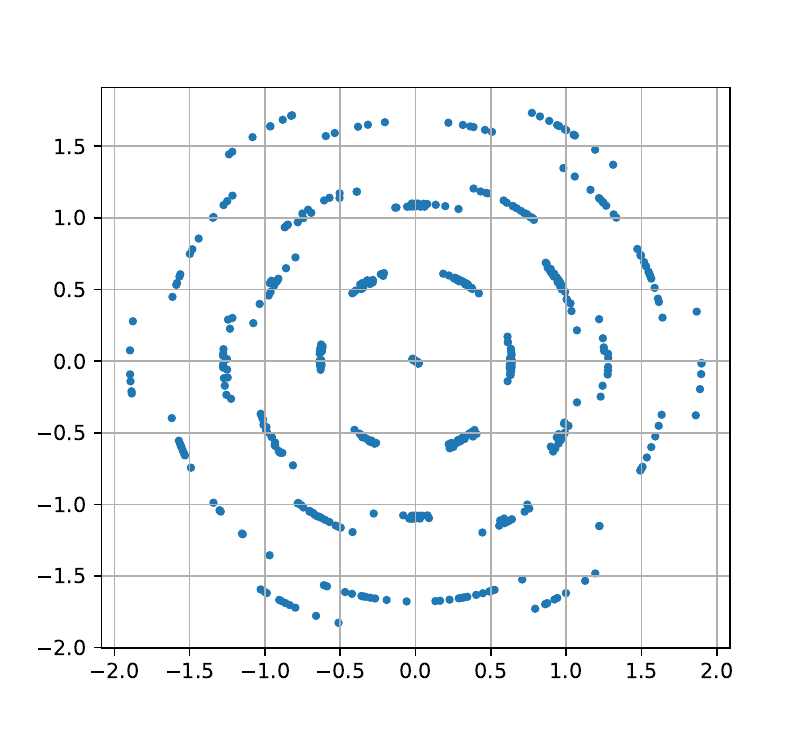}
\end{minipage}%
}%
\subfloat[GAM \cite{GAM}]{
\begin{minipage}[t]{0.42\linewidth}
\centering
\includegraphics[width=1.25in]{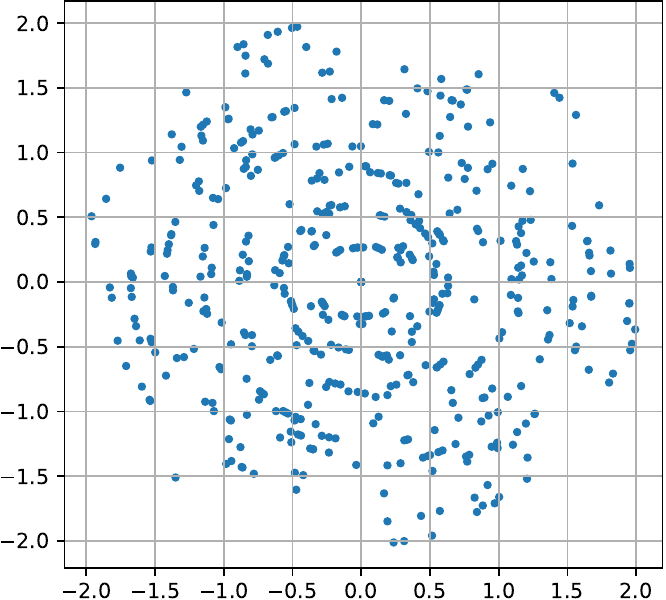}
\end{minipage}%
}%
\centering
\caption{ The effect of PN on different SCMA codebooks with $\sigma_{\mathrm{p}}^{2}=0.01$ and $E_b/N_0=10 \text{dB}$.}
\label{PN_codebook}
\vspace{-1.5em}
\end{figure}

We consider a $K\times J$ SCMA system, where $J$ users communicate over $K$ resource nodes (RNs). The overloading factor is defined as $\lambda =J/K>100\%$. In SCMA, each user is assigned with a unique codebook, denoted by $ \boldsymbol{\mathcal { X }}_{j} \in \mathbb C^{K \times M}$, consisting of $M$ codewords with a dimension of $K$. At the base station (BS) side, each SCMA encoder maps $\log_{2}M$ binary bits to  a length-$K$ sparse complex codeword $\mathbf{x}_{j}=\left[x_{j, 1}, x_{j, 2}, \ldots, x_{j, K}\right]^{T}$ drawn from the codebook $\boldsymbol{\mathcal { X }}_{j}$. Each codeword has only $N$ nonzero entries and the positions of the non-zeros elements remain the same within a codebook. Such a sparse system can efficiently represented by a  factor graph which illustrates  the sharing of the  RNs among multiple  user nodes (UNs). An SCMA factor graph with $K=4$ and $J=6$ is shown in Fig. \ref{Factor}. For a downlink SCMA system, users' data are superimposed at the BS, constituting a superimposed constellation $\mathbf \Phi \in \mathbb C^{K \times M^J}$ with $M^{J}$ different superimposed codewords.   Denote by $\mathbf{w}=\sum_{j=1}^{J} \mathbf{x}_{j} \in \mathbf \Phi  $ the  superimposed codeword. Consider a Gaussian channel,  the received signal affected by unknown PN  can be expressed as  \cite{soft}
\begin{equation}
\small
\label{remodel_pn}
\mathbf{r}=\mathbf{w}e^{j\boldsymbol{\theta}}+\mathbf{n},
\end{equation}
where $\mathbf r \in \mathbb C^{K \times 1} $ denotes the received signal vector,   {$\mathbf{n}=[n_{1},...,n_{K}]^T$ denotes  the complex Gaussian  noise  vector with its element $n_{k} \sim \mathcal{C N}\left(0, N_{0}\right)$, $\boldsymbol{\theta}=[\theta_{1},\theta_{2},\ldots, \theta_{K}]^\mathcal T$ denotes the phase error vector which is subject to a  Gaussian distribution with zero  mean and variance $\sigma_{\mathrm{p}}^{2}$, i.e.,  $\theta _{k}\sim \mathcal{N}\left(0, \sigma_{\mathrm{p}}^{2}\right)$.}  By leveraging  the  codebook sparsity,   MPA  can be employed for low-complexity decoding whose error performance approaches that of maximum likelihood (ML) receiver. {Fig. \ref{PN_codebook} shows an example of the received signal at the first resource node impaired by PN, where  Chen's codebook \cite{Chen} and golden angle modulation (GAM) codebook \cite{GAM}  are considered. When PN is introduced, the constellation points in the original diagram undergo noticeable rotation, and potentially leading to overlap at higher PN levels, which increases error rates.}

In the following, we will discuss the ML decision region in the
presence of PN.  For the communication  model in (\ref{remodel_pn}), the likelihood based on the received signal is given by \cite{likelyhood_kam} 
\begin{equation}
\small
\label{likeli1}
    f\left(\boldsymbol{r} \mid \mathbf{w}\right)=\prod_{k=1}^{K} \int_{-\pi}^{\pi} p\left(r_{k} \mid w_{k}, \theta_{k}\right) p\left(\theta_{k} \mid \overline{\boldsymbol{r}}_{k}, w_{k}\right) \mathrm{d} \theta_{k},
\end{equation}
where $\overline{\boldsymbol{r}}_{k}\triangleq\left[r_{1}, \ldots, r_{k-1}, r_{k+1}, \ldots, r_{K}\right]^{T}$ denotes   the received codeword by excluding the $k$th symbol, and the \textit{a posteriori}  probability density function (pdf) of the PN is given by $p\left(\theta_{k}\mid\overline{\boldsymbol{r}}_{k}, w_{k}\right)=p\left(\theta_{k}\right)=\mathcal{N}\left(0, \sigma_{\mathrm{p}}^{2}\right)$.  Then, the ML decision of the received superimposed codeword is given as 
\begin{equation}
\label{ML1}
    \mathbf{\hat{w}} \triangleq \underset{\mathbf{w}  \in \mathbf \Phi}{\operatorname{argmax}} \; \; \;  f\left(\mathbf{r} \mid \mathbf{w}\right).
\end{equation}
 In the sequel, we will derive approximate expressions for the likelihood (\ref{likeli1}), which in turn allows us to obtain approximate ML detectors. These calculations play a pivotal role in the optimization process when designing new SCMA codebooks. 
 
 \vspace{-0.5em}
\section{Proposed Design Criteria of Sparse Codebook in the Presence of PN}
\label{PEPS}
In this section, we first derive the decision rule of the received codewords with   instantaneous phase error. Then, we present the   proposed codebook design criteria.

\vspace{-1em}
\subsection{ML Decision Rule Based on  Phase Error Approximation}
To derive the ML decision, we first rewrite (\ref{likeli1}) as \cite{soft}
\begin{equation}
\small
\label{likeli2}
\begin{aligned}
  f\left(\boldsymbol{r}  \! \mid  \! \mathbf{w}\right)&=\prod_{k=1}^{K} \int_{-\pi}^{\pi} p\left(r_{k} \mid w_{k}, \theta_{k}\right) p\left(\theta_{k}  \!\mid  \! \overline{\boldsymbol{r}}_{k}, w_{k}\right) \mathrm{d} \theta_{k}\\
& =\prod_{k=1}^{K}\int_{-\pi}^{\pi} p\left(r_{k}, \theta_{k}  \! \mid  \! w_{k}, \overline{\mathbf{r}}_{k}\right) d \theta_{k}   =\prod_{k=1}^{K}p\left(r_{k}  \! \mid \! w_{k}, \overline{\mathbf{r}}_{k}\right).
\end{aligned}
\end{equation}
Before calculating (\ref{likeli2}),  we propose to approximate the received signal $\mathbf r$ as follows:
\begin{equation}
\label{signalmodel}
\begin{aligned}
 r_{k} & =w_{k} e^{j \theta_{k}}+n_{k} \\
& =\left|w_{k}\right| e^{j \arg \left\{w_{k}\right\}} e^{j \theta_{k}}+n_{k}^{\prime} e^{j \arg \left\{w_{k}\right\}} \\
& =\left(\left|w_{k}\right| e^{j \theta_{k}}+n_{k}^{\prime}\right) e^{j \arg \left\{w_{k}\right\}} \\
&  \overset{(\mathrm i)}{\approx}\left(\left|w_{k}\right|+\Re\left\{n_{k}^{\prime}\right\}+j\left(\left|w_{k}\right| \theta_{k}+\Im\left\{n_{k}^{\prime}\right\}\right)\right) e^{j \arg \left\{w_{k}\right\}} ,      
\end{aligned}
\end{equation}
where step $ {(\mathrm i)}$ is obtained by using the approximation  of  $e^{j \theta_{k}} \approx 1+j \theta_{k}$,  $r_{k} $ denotes the $k$th entry of  $\mathbf r$, and $n_{k}^{\prime} \triangleq n_{k} e^{-j \arg \left\{w_{k}\right\}}$.   In general, $ {(\mathrm i)}$ holds tighter as $\sigma_{\mathrm{p}}^{2}$  decreases.  
 Essentially, $r_{k}$
can be expressed as  a 2-tuple $\left[\mu_{k}, \nu_{k}\right]$ comprising of the real and imaginary parts of $r_{k}$, where $\mu_{k} \triangleq \Re\left\{r_{k} e^{-j \arg \left\{w_{k}\right\}}\right\}-\left|w_{k}\right|, \nu_{k} \triangleq \Im\left\{r_{k} e^{-j \arg \left\{w_{k}\right\}}\right\}$. The pdf of this tuple is a bi-variate Gaussian distribution PDF conditioned on $w_{k}$, with mean value given as
\begin{equation}
\label{mean}
    \mathbb{E}\left[\mu_{k}, \nu_{k}\right]=\mathbb{E}\left[\Re\left\{n_{k}^{\prime}\right\},\left|w_{k}\right| \theta_{k}+\Im\left\{n_{k}^{\prime}\right\}\right] 
=[0\quad0]^{T},
\end{equation}
and covariance 
\begin{equation}
\small
\label{variance}
    \begin{array}{l}
\mathbb{E}\left[\begin{array}{cc}\left|\Re\left\{n_{k}^{\prime}\right\}\right|^{2} & \Re\left\{n_{k}^{\prime}\right\}\left(\left|w_{k}\right| \theta_{k}+\Im\left\{n_{k}^{\prime}\right\}\right) \\\Re\left\{n_{k}^{\prime}\right\}\left(\left|w_{k}\right| \theta_{k}+\Im\left\{n_{k}^{\prime}\right\}\right) & || w_{k}\left|\theta_{k}+\Im\left\{n_{k}^{\prime}\right\}\right|^{2}\end{array}\right] \\=\left[\begin{array}{cc}N_{0} / 2 & 0 \\0 & \sigma_{\mathrm{p}}^{2}\left|w_{k}\right|^{2}+N_{0} / 2\end{array}\right] .
\end{array}
\end{equation}

 Using (\ref{mean}) and (\ref{variance}), the conditional pdf of $\mathbf{r}$ for given $\textbf{w}$ is
\begin{equation}
\small
\label{f_pn}
\begin{aligned}
&f\left(\mathbf{r} \mid \mathbf{w}\right)  =\prod_{k=1}^{K}p\left(r_{k} \mid w_{k}, \overline{\mathbf{r}}_{k}\right) \\
& =\frac{e^{-\frac{1}{2}\sum_{k=1}^{K}\left(\frac{\left(\Re\left\{r_{k} e^{-j \arg \left\{w_{k}\right\}}\right\}-\left|w_{k}\right|\right)^{2}}{N_{0} / 2}+\frac{\left(\Im\left\{r_{k} e^{-j \arg \left\{w_{k}\right\}}\right\}\right)^{2}}{\sigma_{\mathrm{p}}^{2}\left|w_{k}\right|^{2}+N_{0} / 2}\right)}}{\left(2 \pi\right)^K \sqrt{\left(N_{0} / 2\right)^K\prod_{k=1}^{K}\left(\sigma_{\mathrm{p}}^{2}\left|w_{k}\right|^{2}+N_{0} / 2\right)}} .
\end{aligned}
\end{equation}
After taking negative logarithm of (\ref{f_pn}),  the ML decision rule is  expressed as
{
\begin{equation}
\small
\label{LPHN}
\begin{aligned}
       \hat{\mathbf{w}}&=\underset{\mathbf{w} \in \mathbf \Phi}{\operatorname{argmax}}\quad f\left(\mathbf{r} \mid \mathbf{w}\right)\\&=\underset{\mathbf{w} \in \mathbf \Phi}{\operatorname{argmin}}\quad \left\{\sum_{k=1}^{K}\frac{\left(\Re\left\{r_{k} e^{-j \arg \left\{w_{k}\right\}}\right\}-\left|w_{k}\right|\right)^{2}}{N_{0} / 2} \right.\\ & \left.\quad +\frac{\left(\Im\left\{r_{k} e^{-j \arg \left\{w_{k}\right\}}\right\}\right)^{2}}{\sigma_{\mathrm{p}}^{2}\left|w_{k}\right|^{2}+N_{0} / 2}+\log \left(\sigma_{\mathrm{p}}^{2}\left|w_{k}\right|^{2}+N_{0} / 2\right) \right\}\\
       &\triangleq \underset{\mathbf{w} \in \mathbf \Phi}{\operatorname{argmin}} \quad L_\mathbf{w} .
\end{aligned}
\end{equation}}

\subsection{PEP  Analysis of SCMA Impaired by PN}
\vspace{-1em}
 Due to PN and  Gaussian noise, the transmitted signal $\mathbf{w} \in \mathbf \Phi $ is assumed to be erroneously decoded to another codeword $\hat{\mathbf{w}} \in \mathbf \Phi, \mathbf w \neq \hat{\mathbf w} $. Denote $\operatorname{Pr}\{\mathbf{w} \rightarrow \hat{\mathbf{w}}\}$ by the PEP between $\mathbf{w}$ and $\hat{\mathbf{w}}$,  then the average PEP, denoted as $P_{e}$,   is  upper-bounded by averaging over all error events, i.e.,
\begin{equation}
\label{Pe}
\small
    P_{e} \leq \frac{1}{M^{J}} \sum_{\mathbf{w}} \sum_{\hat{\mathbf{w}} \neq \mathbf{w}}  \operatorname{Pr}\{\mathbf{w} \rightarrow \hat{\mathbf{w}}\}.
\end{equation}
Moreover,  by leveraging the ML decision rule derived above, we can express the probability of a pairwise error event as follows:
\begin{equation}
\small
    \operatorname{Pr}\{\mathbf{w} \rightarrow \hat{\mathbf{w}}\}=\operatorname{Pr}\left(L_{\mathbf{w}}-L_{\hat{\mathbf{w}}}>0 \mid \mathbf{w}\right).
\end{equation}
The implication of (11) is  that when error decoding occurs, the performance metric $L_{\hat{\mathbf{w}}}$ becomes smaller than  that of $L_{\mathbf{w}}$.  For the likelihood in (\ref{LPHN}), after defining $\boldsymbol{\eta}=L_{\mathbf{w}}-L_{\hat{\mathbf{w}}}$, the difference  between   $L_{w_{k}}$ and $ L_{\hat{w}_{k}}$ is given as
\begin{equation}
\label{eta_k}
\small
\begin{aligned}
   \eta_{k}& =L_{w_{k}}-L_{\hat{w}_{k}}\\
   & =\log \left(\frac{\sigma_{\mathrm{p}}^{2}\left|w_{k}\right|^{2}+N_{0} / 2}{\sigma_{\mathrm{p}}^{2}\left|\hat{w}_{k}\right|^{2}+N_{0} / 2}\right)\\
    & \left.-\underbrace{\frac{\left(\Re\left\{r e^{-j \arg \left\{\hat{w}_{k}\right\}}\right\}-\left|\hat{w}_{k}\right|\right)^{2}}{N_{0} / 2}}_{\triangleq V_{1}}-\underbrace{\frac{\left(\Im\left\{r e^{-j \arg \left\{\hat{w}_{k}\right\}}\right\}\right)^{2}}{\sigma_{\mathrm{p}}^{2}\left|\hat{w}_{k}\right|^{2}+N_{0}/2}}_{\triangleq V_{2}}\right.\\
    &+\frac{\left(\Re\left\{r e^{-j \arg \left\{w_{k}\right\}}\right\}-\left|w_{k}\right|\right)^{2}}{N_{0} / 2} +\frac{\left(\Im\left\{r e^{-j \arg \left\{w_{k}\right\}}\right\}\right)^{2}}{\sigma_{\mathrm{p}}^{2}\left|w_{k}\right|^{2}+N_{0} / 2} .
\end{aligned}
\end{equation}
Given the superimposed codeword $\mathbf{w}$ has been transmitted, the real and imaginary parts of the received signal $\mathbf r$ from (\ref{signalmodel}) are expressed below:
\begin{equation}
\label{ReIm}
\small
\begin{aligned}
     \Re\left\{r_{k} e^{-j \arg \left\{w_{k}\right\}}\right\}&=\left|w_{k}\right|+\Re\left\{n_{k}^{\prime}\right\},\\
     \Im\left\{r_{k} e^{-j \arg \left\{w_{k}\right\}}\right\}&=\left|w_{k}\right| \theta_{k}+\Im\left\{n_{k}^{\prime}\right\},\\
      \Re\left\{r_{k} e^{-j \arg \left\{\hat{w}_{k}\right\}}\right\}&=(\left|w_{k}\right|+\Re\left\{n_{k}^{\prime}\right\})\cos(\Delta _{w})\\&-(\left|w_{k}\right| \theta_{k}+\Im\left\{n_{ k}^{\prime}\right\})\sin(\Delta _{w}),\\
     \Im\left\{r_{k} e^{-j \arg \left\{\hat{w}_{k}\right\}}\right\}&=(\left|w_{k}\right|+\Re\left\{n_{k}^{\prime}\right\})\sin(\Delta _{w})\\& +(\left|w_{k}\right| \theta_{k}+\Im\left\{n_{k}^{\prime}\right\})\cos(\Delta _{w}),
\end{aligned}
\end{equation}
 where $\Delta _{w}=\arg \left\{w_{k}\right\}-\arg \left\{\hat{w}_{k}\right\}$.  Substituting (\ref{ReIm}) into (\ref{eta_k}),  we obtain
\begin{equation}
    \begin{aligned}
{\eta}_{k} =&\log \left(\frac{\sigma_{\mathrm{p}}^{2}\left|w_{k}\right|^{2}+N_{0} / 2}{\sigma_{\mathrm{p}}^{2}\left|\hat{w}_{k}\right|^{2}+N_{0} / 2}\right)+\frac{\left(\Re\left\{n_{k}^{\prime}\right\}\right)^{2}}{N_{0} / 2}-V_{1}-V_{2} \\&+\frac{\left(\left|w_{k}\right|^2 \theta_{k}^2+2\left|w_{k}\right| \theta_{k}\Im\left\{n_{k}^{\prime}\right\}+\Im^2\left\{n_{k}^{\prime}\right\}\right)}{\sigma_{\mathrm{p}}^{2}\left|w_{k}\right|^{2}+N_{0} / 2},
\end{aligned}
\end{equation}
where the exact expressions  of $V_{1}$ and $V_{2}$ can be found in (\ref{eta_k}). The   mean values  of $V_{1}$ and $V_{2}$ 
 are respectively given by
\begin{equation}
\small
E\{V_{1}\}=\frac{(\left|w_{k}\right|\cos\Delta_{w}-\left|\hat{w}_{k}\right|)^2+\left|w_{k}\right|^2\sin^2\Delta_{w}\sigma_{\mathrm{p}}^2+N_{0}/2}{N_{0}/2},
\end{equation}
\begin{equation}
\small
E\{V_{2}\}=\frac{\left|w_{k}\right|^2(\sin^2\Delta_{w}+\sigma_{\mathrm{p}}^2\cos^2\Delta_{w})+N_{0}/2}{\sigma_{\mathrm{p}}^{2}\left|\hat{w}_{k}\right|^{2}+N_{0}/2}.
\end{equation}

Hence, the mean value and  covariance of $\eta_{k}$ conditioned on $w_{k}$ are   respectively given as 
\begin{equation}
\small
\label{eta_mean_var}
\begin{aligned}
E_{\text{PN}}& \{\eta_{k}|w_{k}\rightarrow \hat w_{k}\} \\ & =2+\log \left(\frac{\sigma_{\mathrm{p}}^{2}\left|w_{k}\right|^{2}+N_{0} / 2}{\sigma_{\mathrm{p}}^{2}\left|\hat{w}_{k}\right|^{2}+N_{0} / 2}\right)    -E\{V_{1}\}-E\{V_{2}\}\\
\sigma_{\text{PN}}^2 & \{\eta_{k}|w_{k}\rightarrow  \hat w_{k}\}\\ &=\mathbb{E}\left\{\left(\eta_{k}-\mathbb{E}\left\{\eta_{k} \mid w_{k},  \hat w_{k} \right\}\right)^{2} \mid w_{k}\right\}\\&=\mathbb{E}\left\{L_{w_{k}}^{2} \mid w_{k}\right\}+\mathbb{E}\left\{L_{\hat{w}_{k}}^{2} \mid w_{k}\right\}\\ & \quad \quad -2 \mathbb{E}\left\{L_{w_{k}} L_{\hat{w}_{k}} \mid w_{k}\right\}-\left(\mathbb{E}\left\{\eta_{k} \mid w_{k}\right\}\right)^{2}\\
&=4+2a_{k}^2+4a_{k}b_{k}+2c_{k}^2+4c_{k}d_{k}+4e_{k}f_{k}\\& \quad \quad +4e_{k}g_{k}
-4b_{k}e_{k}-4\sin^2(\Delta_{w})(h_{k}+\frac{1}{i_{k}})\\&\quad \quad -4\cos^2(\Delta_{w})(1+\frac{h_{k}}{i_{k}}),
\end{aligned}
\end{equation}
where
\begin{equation}
\small
\label{abcde}
\begin{aligned}
&a_{k}=\frac{\left|w_{k}\right|^{2}\sin^2{\Delta_{w}}\sigma_{\mathrm{p}}^{2}+N_{0} / 2}{N_{0} / 2},
b_{k}=\frac{(\left|w_{k}\right|\cos\Delta_{w}-\left|\hat{w}_{k}\right|)^2}{N_{0} / 2},\\
&c_{k}=\frac{\left|w_{k}\right|^{2}\cos^2{\Delta_{w}}\sigma_{\mathrm{p}}^{2}+N_{0} / 2}{\sigma_{\mathrm{p}}^{2}\left|\hat{w}_{k}\right|^2+N_{0} / 2},
d_{k}=\frac{\left|w_{k}\right|^{2}\sin^2{\Delta_{w}}}{\sigma_{\mathrm{p}}^{2}\left|\hat{w}_{k}\right|^2+N_{0} / 2}\\
&e_{k}=\frac{\left|w_{k}\right|^{2}\sin^2{\Delta_{w}}\sigma_{\mathrm{p}}^{2}}{\sigma_{\mathrm{p}}^{2}\left|\hat{w}_{k}\right|^2+N_{0} / 2},
f_{k}=\frac{\left|w_{k}\right|^{2}\cos^2{\Delta_{w}}\sigma_{\mathrm{p}}^{2}}{N_{0} / 2},\\
&g_{k}=\frac{\left|\hat{w}_{k}\right|^2-\left|w_{k}\right|^{2}\cos^2{\Delta_{w}}}{N_{0} / 2},
h_{k}=\frac{\sigma_{\mathrm{p}}^{2}\left|w_{k}\right|^2+N_{0} / 2}{N_{0} / 2},\\
&i_{k}=\frac{\sigma_{\mathrm{p}}^{2}\left|\hat{w}_{k}\right|^2+N_{0} / 2}{N_{0} / 2}.
\end{aligned}
\end{equation}
With  (\ref{eta_mean_var}),  the probability of a pairwise error event can be obtained as
\begin{equation}
\label{Pr}
\small
    \operatorname{Pr}\left(\boldsymbol{\eta}>0 \mid \mathbf{w}\rightarrow \mathbf{\hat{w}}\right)\approx \mathcal{Q}\left( \frac{- E_{\text{PN}}\{\boldsymbol{\eta}|\mathbf{w}\rightarrow \mathbf{\hat{w}}\}}{\sqrt{\sigma_{\text{PN}}^2\{\boldsymbol{\eta}|\mathbf{w}\rightarrow \mathbf{\hat{w}}\}} } \right),
\end{equation}
where $ E_{\text{PN}}\{\boldsymbol{\eta}|\mathbf{w}\rightarrow \mathbf{\hat{w}}\}=\sum_{k=1}^{K}E_{\text{PN}}\{\eta_{k}|w_{k}\rightarrow \hat{w}_{k}\}$ and $\sigma_{\text{PN}}^2\{\boldsymbol{\eta}|\mathbf{w}\rightarrow \mathbf{\hat{w}}\}=\sum_{k=1}^{K}\sigma_{\text{PN}}^2\{\eta_{k}|w_{k}\rightarrow \hat{w}_{k}\}$, and $  \mathcal{Q}(t)=\frac{1}{\pi}\int _{0}^{+\infty } \frac{e^{ -\frac{x^2 }{2}(t^2+1)}  }{t^2+1} \text{d}t 
$ is the Gaussian $  \mathcal{Q} $-function.
Substituting (\ref{Pr}) into (\ref{Pe}), the average PEP $P_{e}$ is upper bounded by
\begin{equation}
\small
\label{PeLPHN}
\begin{aligned}
    P_{e} &\leq \frac{1}{M^{J}} \sum_{\mathbf{w}} \sum_{\hat{\mathbf{w}} \neq \mathbf{w}}\mathcal{Q}\left( \frac{- E_{\text{PN}}\{\boldsymbol{\eta}|\mathbf{w}\rightarrow \mathbf{\hat{w}}\}}{\sqrt{\sigma_{\text{PN}}^2\{\boldsymbol{\eta}|\mathbf{w}\rightarrow \mathbf{\hat{w}}\}} } \right).
\end{aligned}
\end{equation}
 It should be noted that the argument of the $\mathcal{Q}$ function in (\ref{PeLPHN}) can be used to determine the nearest codewords in the presence of PN.  To construct the optimal codebook, it is desirable to maximize the  following metric:
\begin{equation}
\label{MPND}
\small
\begin{aligned}
    \mathrm{MPNM} =   \min \left \{\frac{- E_{\text{PN}}\{\boldsymbol{\eta}|\mathbf{w}\rightarrow \mathbf{\hat{w}}\}}{\sqrt{\sigma_{\text{PN}}^2\{\boldsymbol{\eta}|\mathbf{w}\rightarrow \mathbf{\hat{w}}\}} }   \vert  \quad \forall \mathbf w , \mathbf{\hat{w}}\in \mathbf \Phi,  \mathbf w \neq \mathbf{\hat{w}}\right\}.
\end{aligned}
\end{equation}
which is  referred to as the  minimum PN metric (MPNM).
 


\color{black}

\section{ Proposed PN Resilient Codebook Design}
\label{CBd}

This section presents PN resilient  sparse codebook design  based on the proposed  MPNM metric.  The  design steps mainly include: the generation of one-dimensional basic constellation, the construction of the $N$-dimension MC and the  optimization of multiple sparse codebooks by maximizing MPNM. 

 \subsection{LP Inspired One-dimensional Constellation Design}

Based on Fig.  \ref{PN_codebook}, one can observe that the  superimposed constellation can be well distinguished from the PN if it has less unique constellation points.  This can be achieved by allowing certain overlapping  within the same dimension.  Due to the multidimensional nature, two transmitted codewords   with overlapped constellation elements  can be distinctly separated in other dimensions.  Such codebook is  called LP codebook in \cite{LPSCMA}, and the proposed MC is called  LP-MC.  Motivated by this feature, a new PAM-based MC, called LP-PAM, is proposed for suppressing  PN.  The length-$M$ one-dimensional LP-PAM constellation, denoted by $\boldsymbol{\mathcal{C}}_{M}$, is composed of $T$ distinct elements and $M-T$ overlapped points. Specifically, the construction of length $T$ vector $\boldsymbol{\mathcal{C}}_{T}$ is summarized as followed:  for even $T$, generate $T$ PAM points; for odd  $T$, generate $T-1$ PAM points and add the original point  to $\boldsymbol{\mathcal{C}}_{T}$.  For example,  for even $T$,  the one-dimensional PAM constellation $\boldsymbol{\mathcal{C}}_T$ can be expressed as 
\begin{equation}
\small
\begin{aligned}
    \boldsymbol{\mathcal{C}}_{T}&=\left[z_{1,1}, z_{1,2}, \ldots, z_{1, T}\right]\\
    &=\left[-r_{T / 2},-r_{T / 2-1}, \ldots,-r_{1}, r_{1}, \ldots, r_{T / 2}\right],
\end{aligned}
\end{equation}
and $\alpha_{m}=r_{m+1} / r_{1}\in(1,+\infty), \quad m=1,2, \ldots, T / 2-1$.  Then, to obtain $\boldsymbol{\mathcal{C}}_{M}$, $M-T$ points in $\boldsymbol{\mathcal{C}}_{T}$ must be overlapped, where the following two  rules are applied to guide the selection of overlapped constellations: {1) The constellation points with lower energy  are
preferentially overlapped to construct $\boldsymbol{\mathcal{C}}_M$, such that  
$\boldsymbol{\mathcal{C}}_\mathrm{MC}$ has a small average energy;  2) The overlapped constellation points should also exhibit certain symmetry, such that  $\boldsymbol{\mathcal{C}}_M$ is also symmetrical and with a zero mean. }


\vspace{-1em}
\subsection{Construction of the $N$-dimension MC}
\begin{figure}
\centering
\subfloat[First dimension]{
\begin{minipage}[t]{1\linewidth}
\centering
\includegraphics[width=3.5in]{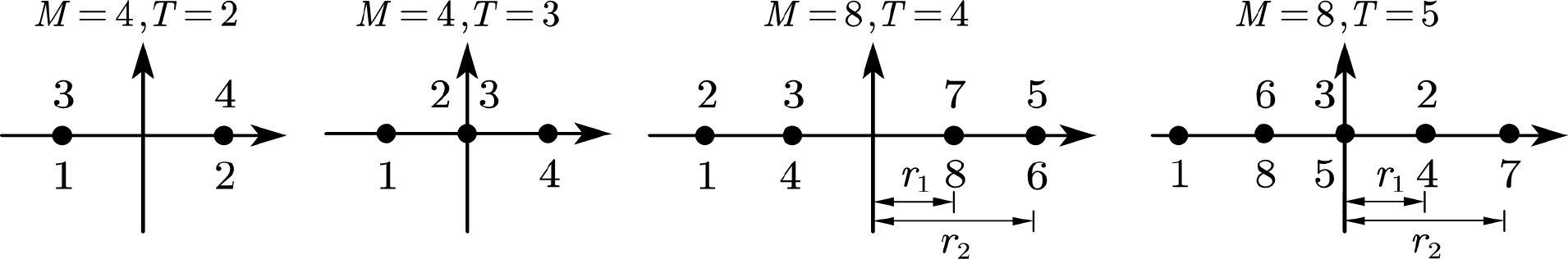}
\end{minipage}%
}%
\quad
\subfloat[Second dimension]{
\begin{minipage}[t]{1\linewidth}
\centering
\includegraphics[width=3.5in]{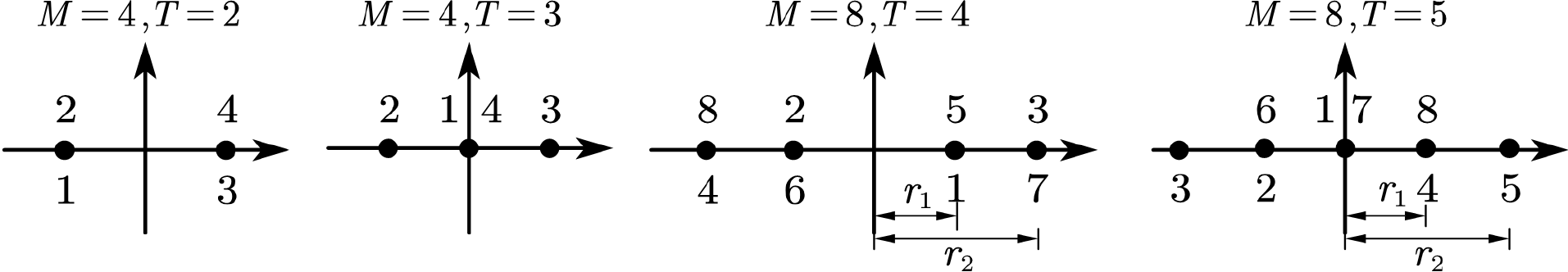}
\end{minipage}%
}%
\centering
\caption{ Example of $\boldsymbol{\mathcal{C}}_{M}$ and $\pi_{2}^{\mathcal{T}}\left(\boldsymbol{\mathcal{C}}_{M}\right)$.}
\label{constellation_CM}
\vspace{-2em}
\end{figure}

After obtaining the one-dimensional constellation $  \boldsymbol {\mathcal{C}}_{M}$, the remaining $N-1$ dimensions may be obtained by the permutation of $\boldsymbol{\mathcal{C}}_{M}$. Let ${\pi}_{n} $ denote the permutation mapping of the $n$th dimension. Then the $N$-dimensional  $\boldsymbol{\mathcal{C}}_{\text{MC}}\in \mathbb{C}^{N \times M}$ can be obtained as
\begin{equation}
\small
\boldsymbol{\mathcal{C}}_{\text{MC}}=\left[\pi_{1}^{\mathcal{T}}\left(\boldsymbol{\mathcal{C}}_{M}\right), \pi_{2}^{\mathcal{T}}\left(\boldsymbol{\mathcal{C}}_{M}\right), \ldots, \pi_{N}^{\mathcal{T}}\left(\boldsymbol{\mathcal{C}}_{M}\right)\right]^{\mathcal{T}}.
\end{equation}
The goal is to find the  $N$ permutations $\pi_{n}, n = 1,2,\cdots,N$  for maximizing the minimum Euclidean distance of  $\boldsymbol{\mathcal{C}}_{\text{MC}}$. {In this paper,  the minimum Euclidean distance (MED) is considered as permutation metric and  the the binary switching algorithm is employed to find the $N$ permutations \cite{LPSCMA}.}  Next, we give an example of the proposed  $\boldsymbol{\mathcal{C}}_{M}$ and $\pi_{2}^{\mathcal{T}}\left(\boldsymbol{\mathcal{C}}_{M}\right)$ in Fig. \ref{constellation_CM}, where $\alpha_{1} =r_{2}/r_{1}$ is a parameter to be optimized.

\vspace{-1em}
\subsection{ Optimization of  Sparse Codebooks}
Upon obtaining  $\boldsymbol{\mathcal{C}}_{\text{MC}}$, phase rotations and energy scaling are applied  to generate multiple sparse codebooks in order to enhance  the  distance properties of the superimposed codewords \cite{LPSCMA}.  Specifically, user $j$’s sparse codebook is designed as $\boldsymbol{\mathcal{C}}_{j}=\mathbf{E}_{j} \mathbf{R}_{j} \boldsymbol{\mathcal{C}}_{\text{MC}}$,  where $\mathbf{R}_{j}$ and $\mathbf{E}_{j} $ denote the phase rotation matrix and power scaling matrix of user $j$, respectively. For simplicity, we define a constellation operator matrix as $\boldsymbol{\Psi}_{j}=\mathbf{E}_{j} \mathbf{R}_{j}$. For a $\boldsymbol{\mathcal{C}}_{\text{MC}}$ with $N = 2$, $\mathbf{E}_{j}$ and $ \mathbf{R}_{j}$ can be expressed as
\begin{equation}
\small
    \mathbf{E}_{j}=\left[\begin{array}{cc}E_{1} & 0 \\0 & E_{2}\end{array}\right], \mathbf{R}_{j}=\left[\begin{array}{cc}e^{j \theta_{1}} & 0 \\0 & e^{j \theta_{2}}\end{array}\right], \boldsymbol{\Psi}_{j}=\left[\begin{array}{cc}\psi_{1} & 0 \\0 & \psi_{2}\end{array}\right]
\end{equation}
where $\psi_{i}=E_{i}e^{j \theta_{i}}, \forall E_{i}>0,1 \leq i \leq N$.  Then, user $j$'s codebook  can be generated by $\boldsymbol{\mathcal { X }}_{j}=\mathbf{V}_{j} \boldsymbol{\Psi}_{j} \boldsymbol{\mathcal{C}}_{\text{MC}}$, where $\mathbf{V}_{j}$ is the binary mapping matrix that maps an $N$-dimensional dense constellation to a $K$-dimensional sparse codebook. We further combine the constellation operation matrix $\mathbf{\Psi}_{j}$ and mapping matrix $\mathbf{V}_{j}$ together, i.e., $\boldsymbol{\mathcal V}_{j}=\mathbf{V}_{j} \boldsymbol{\Psi}_{j} $. In this paper, for the SCMA system  given in  Fig. \ref{Factor},  the   mapping matrices are designed as
	\begin{equation}
 \small
		\begin{array}{@{\hspace{1mm}}c@{\hspace{1mm}}c@{\hspace{1mm}}c@{\hspace{1mm}}}
			\boldsymbol{\mathcal{V}}_1\!=\!\begin{bmatrix}
				\psi_1&0\\
				0&0\\
				0&\psi_3\\
				0&0\\
			\end{bmatrix}, &
			\boldsymbol{\mathcal{V}}_2\!=\!\begin{bmatrix}
				0&0\\
				\psi_1&0\\
				0&0\\
				0&\psi_3\\
			\end{bmatrix}, &
			\boldsymbol{\mathcal{V}}_3\!=\!\begin{bmatrix}
				\psi_2&0\\
				0&\psi_2\\
				0&0\\
				0&0\\
			\end{bmatrix},\\ 
			\boldsymbol{\mathcal{V}}_4\!=\!\begin{bmatrix}
				0&0\\
				0&0\\
				\psi_2&0\\
				0&\psi_2\\
			\end{bmatrix}, &
			\boldsymbol{\mathcal{V}}_5\!=\!\begin{bmatrix}
				\psi_3&0\\
				0&0\\
				0&0\\
				0&\psi_1\\
			\end{bmatrix}, &
			\boldsymbol{\mathcal{V}}_6\!=\!\begin{bmatrix}
				0&0\\
				\psi_3&0\\
				0&\psi_1\\
				0&0\\
			\end{bmatrix}. \\
		\end{array}
	\end{equation}

Hence, based on the proposed multi-dimensional codebook construction scheme, the design of sparse codebook  can be formulated as
\begin{equation}
\small
    \begin{aligned}\mathcal{P}: & \max _{\mathbf{E}, \boldsymbol{\theta}, \boldsymbol{\alpha} } \quad \mathrm{MPNM}(\mathcal{X}) \\
    \text { Subject to } \quad & \sum_{i=1}^{d_{f}} E_{i}=\frac{M J}{K}, E_{i}>0, \\& 0 \leq \theta_{i} \leq \pi, \forall i=1,2, \ldots, d_{f},\\
    &\alpha_{m} \ge 1, m=1,2, \ldots, T / 2-1.
    \end{aligned}
    \label{CBdesign}
\end{equation}
{where rotation angles  $\boldsymbol{\theta}=\left[\theta_{1}, \theta_{2}, \ldots, \theta_{d_{f}}\right]^{\mathrm{T}}$, energy factors  $\mathbf{E}=\left[E_{1}, E_{2}, \ldots, E_{d_{f}}\right]^{\mathrm{T}}$ and scattering factors $\boldsymbol{\alpha}=\left[\alpha_1, \alpha_2,\ldots,\alpha_{T/2-1}\right]^{\mathrm{T}}$ are the parameters to be optimized.} Unfortunately, the optimization in (26) is difficult to
solve due to the non-convex objective function.  Similar to existing constellation and SCMA codebook design  works [7], [9], [13], we solve (26)  with a numerical global search method by using the MATLAB Global Optimization Toolbox.



\begin{figure*}[tbp]
\centering
\begin{subfigure}{0.32\textwidth}
  \includegraphics[width=1\textwidth]{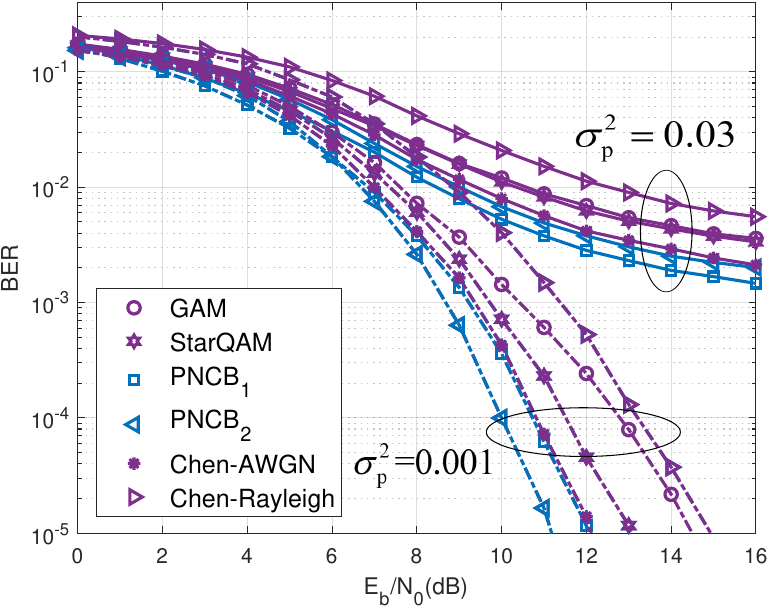}
\caption{ $M=4, \lambda=150\%$.  }
\label{M=4}
\end{subfigure}
\begin{subfigure}{0.32\textwidth}
\includegraphics[width=1\textwidth]{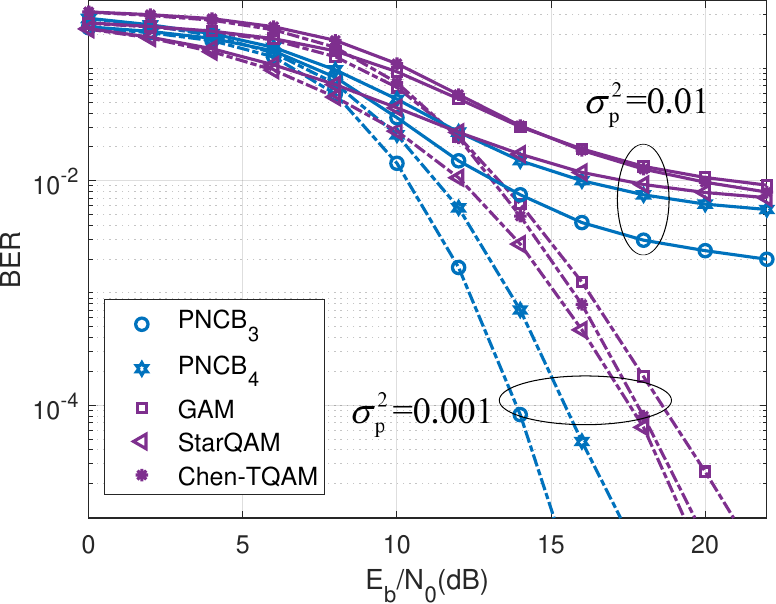}
 \caption{$M=8, \lambda=150\%$. }
\label{M=8}
\end{subfigure}
\begin{subfigure}{0.32\textwidth}
\includegraphics[width=1 \textwidth]{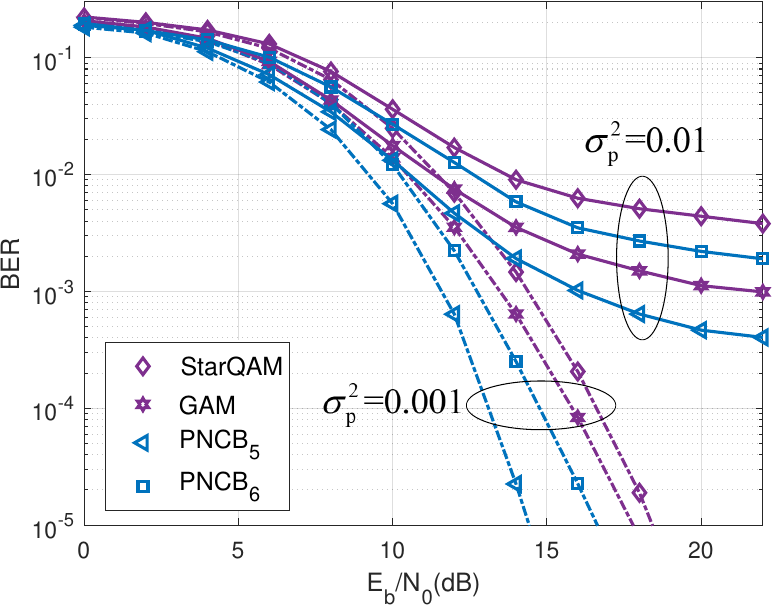}
 \caption{ $M=4, \lambda=200\%$.}
\label{5x10}
\end{subfigure}
 \caption{BER performance comparison  of different codebooks under different PN levels.}
\vspace{-1em}
\end{figure*}

\section{Numerical Results}
\label{Sim}
This section  conducts numerical evaluations for the proposed PNCBs.  In addition to the  SCMA system with $K=4, J=6$, we also consider an  SCMA system with larger size, i.e., $K=5, J=10$, where the mapping matrices in \cite{LPSCMA} is employed.  Given that the proposed criteria  are designed for  high $E_{b}/N_{0}$ regimes, we choose $ E_{b}/N_{0} = 14 $ dB for optimization.  The PNCBs designed with $T=2$ and $3$ for $M=4, \lambda=150\%$ are named as $\mathrm{PNCB}_1$ and $\mathrm{PNCB}_2$, respectively.  The designed  PNCBs with    $T=4$ and $5$ for $M=8, \lambda=150\%$ are named as $\mathrm{PNCB}_3$ and $\mathrm{PNCB}_4$, respectively, and the PNCBs with $T=2$ and $3$ for $M=4, \lambda=200\%$ are named as $\mathrm{PNCB}_5$ and $\mathrm{PNCB}_6$, respectively.   For comparison, we consider Chen's codebook \cite{Chen}, GAM codebook\cite{GAM} and StarQAM codebook\cite{StarQAM}. {Specifically, the codebooks designed in \cite{Chen} for AWGN and Rayleigh fading channels with $M=4$ are referred to as Chen-AWGN and Chen-Rayleigh codebooks, respectively, and the TQAM codebook with $M=8$ is also considered.}

\begin{table}[]
\footnotesize
\begin{center}
\caption{{MPNM and MED values of different codebooks.}}
\label{MPND_table}
\begin{tabular}{c|c|c|c|c}
\hline
System setting        & $\sigma_{\mathrm{p}}^{2}$                    & Codebook    & MPNM &MED \\ \hline
\multirow{6}{*}{\begin{tabular}[c]{@{}l@{}}$M=4, \lambda=150\% $\\ $\mathrm{E_{b}/N_{0}}=10 \text{ dB}$ \end{tabular}} & \multirow{6}{*}{0.03}  & StarQAM     &  1.45 &  0.9 \\ \cline{3-5} 
                      &                        & GAM         &  1.31 & 0.57  \\ \cline{3-5} 
                      &                        & Chen-AWGN      &   1.68  & 1.1 \\ \cline{3-5} 
                      &                        & Chen-Rayleigh      &  1.25   &  0.65 \\ \cline{3-5} 
                      &                        & \textcolor{blue}{$\mathrm{PNCB}_{1}$} &  \textcolor{blue}{1.85}  & 1.06 \\ \cline{3-5} 
                      &                        & \textcolor{blue}{$\mathrm{PNCB}_{2}$} &  \textcolor{blue}{1.70}  & 1.19 \\ \cline{1-5} 
\multirow{5}{*}{\begin{tabular}[c]{@{}l@{}}$M=8, \lambda=150\% $\\ $\mathrm{E_{b}/N_{0}}=15  \text{ dB}$ \end{tabular}} & \multirow{5}{*}{0.01}  & StarQAM     &   3.40 & 0.45 \\ \cline{3-5} 
                      &                        & GAM         &  2.82  &  0.47   \\ \cline{3-5}  
                      &                        & Chen-TQAM         & 3.32  &   0.49   \\ \cline{3-5}
                      &                        & \textcolor{blue}{$\mathrm{PNCB}_{3}$}  & \textcolor{blue}{3.95}   & 0.64  \\ \cline{3-5} 
                      &                        & \textcolor{blue}{$\mathrm{PNCB}_{4}$}  &  \textcolor{blue}{3.82}  &  0.51 \\ \cline{1-5} 
\multirow{4}{*}{\begin{tabular}[c]{@{}l@{}}$M=4, \lambda=200\% $\\ $\mathrm{E_{b}/N_{0}}=15  \text{ dB}$ \end{tabular}} & \multirow{4}{*}{0.01}  & StarQAM     & 2.65  &  0.48 \\ \cline{3-5} 
                      &                        & GAM         & 4.4  &  0.43 \\ \cline{3-5} 
                      &                        & \textcolor{blue}{$\mathrm{PNCB}_{5}$} & \textcolor{blue}{5.1}  & 0.88 \\ \cline{3-5} 
                      &                        & \textcolor{blue}{$\mathrm{PNCB}_{6}$} &   \textcolor{blue}{3.38} & 0.59 \\ \cline{3-5} 
                     \cline{1-5} 
                       \hline
\end{tabular}
\end{center}
\vspace{-2em}
\end{table}

 {
Table \ref{MPND_table} compares the MPNM and MED values of the proposed PNCBs with Chen\cite{Chen}, GAM\cite{GAM} and StarQAM\cite{StarQAM} codebooks for different systems. It is shown that the MPNMs of the proposed codebooks are larger than that of the  other codebooks.}  It should be noted that the codebook with a larger MPNM value enjoy improved  BER performance, and  the MPNM values in Table \ref{MPND_table}  are consistent  with the BER performance  shown in Fig. 4.

Fig. \ref{M=4} compares the BER performance of different codebooks for $M=4, \lambda=150\%$ under different PN levels. {
For  $\sigma_{\mathrm{p}}^{2}=0.001$, the proposed    $\mathrm{PNCB}_1$ and $\mathrm{PNCB}_2$ respectively achieve  about $0.7$ and $1.5$ dB gains over StarQAM at $\mathrm{BER}=10^{-4}$ and $\mathrm{PNCB}_1$ performs slightly better than Chen-AWGN.}  However, for a large PN with   $\sigma_{\mathrm{p}}^{2}=0.03$, the proposed $\mathrm{PNCB}_2$ achieves the best BER performance among all  codebooks.
Fig. \ref{M=8} and Fig. \ref{5x10} compare the BER performance of different codebooks for $M=8, \lambda=150\%$ and $M=4, \lambda=200\%$ under different PN levels,  respectively. As can be seen from   Fig. \ref{M=8}, the proposed  $\mathrm{PNCB}_3$ and $\mathrm{PNCB}_4$ achieve better performance than other codebooks for both $\sigma_{\mathrm{p}}^{2}=0.01$ and $\sigma_{\mathrm{p}}^{2}=0.001$. In Fig. \ref{5x10},  the gain of the proposed $\mathrm{PNCB}_5$ is more prominent, with about $2 $ and $5$ dB gains over GAM at the $\mathrm{BER}=10^{-3}$ for  $\sigma_{\mathrm{p}}^{2}=0.001$  and $\sigma_{\mathrm{p}}^{2}=0.01$, respectively.


 \section{Conclusion}
\label{conclu}
This paper is devoted to PN resilient SCMA codebook design in downlink channels.  First, a novel distance metric named as MPNM has been devised by investigating  the PEP of SCMA impaired by PN. Then, an LP-PAM-based MC has been proposed, followed  by the sparse codebook construction based on the LP-PAM. In addition, the proposed PNCBs are obtained by  maximizing the proposed MPNM. 
 Finally, numerical results have been conducted to demonstrate the superior error performance under PN impairments  for  both $K=4, J=6$ and $K=5, J=10$  SCMA systems.

\vspace{-0.5em}
\bibliography{ref} 
\bibliographystyle{IEEEtran}

\newpage

 




\vfill

\end{document}